\documentclass[sigplan,10pt]{acmart}
\renewcommand\footnotetextcopyrightpermission[1]{}
\usepackage{tikz}
\usetikzlibrary{arrows,calc,fit,backgrounds,positioning}
\usepackage{booktabs} 
\usepackage{style/macros}

\acmYear{2019}
\copyrightyear{2016=9}

\newcommand{\lsem}{\mbox{$\lbrack\!\lbrack$}}
\newcommand{\rsem}{\mbox{$\rbrack\!\rbrack$}}
\newcommand{\sem}[1]{\lsem #1 \rsem}
\newcommand{\op}[1]{\textsc{#1}}
\newcommand{\grad}[1]{#1^{\mathit{grad}}}

\newcommand{\tok}[2]{\langle #1, #2 \rangle}

\begin{document}

\title[Recursive Function Definitions in Static Dataflow Graphs]{%
  Recursive Function Definitions in Static Dataflow Graphs and their
  Implementation in TensorFlow}

\author{Kelly Kostopoulou}
\affiliation{%
  \institution{Columbia University}
}
\email{kelkost@cs.columbia.edu}
\author{Angelos Charalambidis}
\orcid{0000-0001-7437-410X}
\affiliation{%
  \institution{Harokopio University of Athens}
}
\email{acharal@hua.gr}
\author{Panos Rondogiannis}
\affiliation{%
  \institution{University of Athens}
}
\email{prondo@di.uoa.gr}

\renewcommand{\shortauthors}{K. Kostopoulou et al.}

\begin{abstract}
Modern machine learning systems represent their computations as dataflow graphs.
The increasingly complex neural network architectures crave for more powerful
yet efficient programming abstractions. In this paper we propose an efficient
technique for supporting recursive function definitions in dataflow-based
systems such as TensorFlow. The proposed approach transforms the given recursive
definitions into a static dataflow graph that is enriched with two simple yet
powerful dataflow operations. Since static graphs do not change during
execution, they can be easily partitioned and executed efficiently in
distributed and heterogeneous environments. The proposed technique makes heavy
use of the idea of \emph{tagging}, which was one of the cornerstones of dataflow
systems since their inception. We demonstrate that our technique is compatible
with the idea of \emph{automatic differentiation}, a notion that is crucial for
dataflow systems that focus on deep learning applications. We describe the
principles of an actual implementation of the technique in the TensorFlow
framework, and present experimental results that demonstrate that the use of
tagging is of paramount importance for developing efficient high-level
abstractions for modern dataflow systems.
\end{abstract}

\ccsdesc[500]{Software and its engineering~Data flow languages}
\ccsdesc[500]{Software and its engineering~Recursion}
\ccsdesc[300]{Software and its engineering~Distributed programming languages}
\ccsdesc[300]{Software and its engineering~Control structures}
\ccsdesc[100]{Computing methodologies~Neural networks}

\keywords{Tagged dataflow, recursive functions, TensorFlow}

\settopmatter{printfolios=true}
\maketitle

\section{Introduction}
\label{intro}

The interest in the dataflow model has been recently renewed by the latest
developments in scalable machine learning systems. Modern dataflow systems
express computations as dataflow graphs of processing nodes that can work
independently and in parallel. In such a formalism, the data dependencies
between the nodes are explicit and consequently individual nodes can be
distributed in multiple machines and hardware accelerators. These
characteristics render the dataflow model highly desirable for expressing
scalable machine learning tasks.

The popularity of deep learning and its recent successes in solving challenging
problems in various domains, lead to proposals of more complex neural network
architectures. The Recurrent Neural Networks (RNNs)~\cite{lstm} is a prominent
example that uses an elaborate architecture in order to exhibit temporal dynamic
behavior. RNNs benefit from the ability of the machine learning framework to
efficiently implement control-flow decisions, such as conditionals and
iteration, in terms of dataflow graphs. However, more complex neural network
architectures require dataflow systems that can support more demanding
control-flow, such as for example recursion.

\subsection{Implementations of Control Flow}
Current dataflow frameworks implement iteration and recursion using two main
approaches. In the simpler approach, usually coined as the ``out-of-graph''
approach~\cite{chainer,pytorch}, the program is expressed in a client
programming language such as Python and the framework exploits the control-flow
support of the host language. Even though this approach admits a relatively
simple implementation, it requires a continuous interaction between the client
and the dataflow system, resulting in reduced performance. Moreover, in this
approach the dataflow graphs are usually small parts of the whole program, and
the optimizations that can be applied are limited. In order to remedy these
limitations, some machine learning frameworks employ the ``in-graph'' approach
in which the control-flow operations are expressed directly in the dataflow
graph. In this approach the dataflow execution engine is responsible for
implementing the control-flow decisions.

There are two distinct proposals to implement control-flow constructs in the
``in-graph'' approach. The first, relatively straightforward, way suggests that
the dataflow graph is allowed to change at runtime. In this approach the
control-flow operations are represented as nodes in the dataflow graph and when
executed they effectively replace themselves with new nodes of the graph. In
this way one can simulate a dynamic unfolding of the graph. The second, more
elaborate way is to assume that the dataflow graph is fixed, ie., it can not
change at runtime. As it turns out, this \emph{fixed-graph} approach is
preferred by the established deep learning frameworks such as
TensorFlow~\cite{tensorflow_iteration}. There are good reasons for choosing the
fixed-graph idea. For example, an immediate consequence of knowing the graph
ahead of time is that aggressive optimizations and distributed device placement
algorithms can be performed at compile time. Therefore, it is highly desirable,
whenever possible, to express computations as a fixed dataflow graph.

However, it is not always apparent how iteration or recursion can be expressed
in a dataflow graph that is not allowed to expand at runtime. The main concern
that immediately arises is that the nodes of the fixed graph must be reactivated
with different inputs and in different points in time during the execution. In
order to not mix different reactivations of the nodes, the simple dataflow model
has to be extended with the notion of \emph{tags}, i.e., labels that travel with
the data and essentially indicate the {\em context} that the data should be
evaluated under. Additionally, these tagged dataflow graphs must include
``special'' operators that can appropriately manipulate such tags.
TensorFlow~\cite{tensorflow} implements iteration using the aforementioned
tagging technique~\cite{tensorflow_iteration}. However, when it comes to
recursion, TensorFlow resorts to using expandable dataflow graphs, a choice
which diminishes the advantages of the fixed-graph approach.

\subsection{Contributions}
In this paper we consider the problem of efficiently implementing recursive
function definitions using dataflow graphs that remain fixed at runtime. We
assume that the initial program is a set of dataflow graphs each representing a
different function and we allow nodes that correspond to a defined function to
occur as nodes in the graphs. We base our work on the \emph{tagged dataflow}
model~\cite{WatsonG79,ArvindC83,mit-dataflow}. The initial program is
transformed into a single graph by replacing the operators with the graphs that
represent functions. We also use tags to distinguish between different function
calls. The tags, in our proposal, are lists of labels and each label indicates a
different call of a function. We demonstrate that our proposal can relatively
easily be embedded in TensorFlow's runtime, and we demonstrate that the
resulting system is efficient and appropriate for deep learning applications.
The contributions of the paper can be summarized as follows:
\begin{itemize}
\item We devise a purely tagged approach for implementing recursive functions in
      a dataflow environment. Our approach transforms a set of dynamic (ie.,
      expandable) dataflow graphs, into a single static (ie., non-expandable)
      graph which uses two simple (yet powerful) dataflow operations. As a
      result, our technique ensures that dataflow graphs remain fixed during
      runtime. The conceptual origins of our work can be traced back to
      techniques that were developed in the 80s,
      namely~\cite{WatsonG79,ArvindC83,mit-dataflow,manchester-dataflow,yaghi,jfp1},
      but our approach has distinguishing characteristics from each one of them.
      A detailed comparison with the aforementioned techniques is given in
      Section~\ref{related-work}.

\item We demonstrate that our technique is compatible with the idea of
      \emph{automatic differentiation}~\cite{autodiff-survey}, a notion that is
      crucial for dataflow systems that focus on deep learning applications. In
      particular we demonstrate how we can efficiently support automatic
      differentiation in recursive dataflow graphs that have been obtained using
      the proposed tagging approach.

\item We describe an actual implementation of our approach under the TensorFlow
      framework. In particular, we discuss the extensions that are required to
      the computational model of TensorFlow in order to execute dataflow graphs
      that contain the new dataflow operators required for handling recursion.
      The most demanding changes to the computational model concern the
      distributed execution of these static dataflow graphs.

\item We present experimental evidence which confirms that the tagged execution
      outperforms the dynamic (ie., graph expanding) execution of recursion in
      TensorFlow. Moreover, we present experiments regarding the
      efficiency of automatic differentiation which is a key component of deep
      learning applications.
\end{itemize}
As an overall remark, our work suggests that the extensive use of {\em tags} is
a promising direction that current implementations of dataflow systems should
follow. Tags offer increased parallelism and may prove to offer viable solutions
in cases where current technologies appear to bottleneck.

\subsection{Structure of the paper}
The structure of the remaining part of the paper has as follows.
Section~\ref{tagged-dataflow} presents background on tagged-dataflow; most of
the ideas in this section can be traced back to the early steps of dataflow
systems. Section~\ref{transformation} presents the proposed transformation
algorithm from dynamic dataflow graphs to static ones that contain two new
dataflow operations. Section~\ref{autodiff} presents how automatic
differentiation can be implemented efficiently in the static graphs that result
from our transformation. Section~\ref{impl} describes an implementation of the
proposed transformation in the TensorFlow framework. Section~\ref{evaluation}
presents a performance evaluation of our implementation. Finally,
Section~\ref{related-work} compares our approach with related work and
Section~\ref{conclusions} concludes the paper giving pointers to future work.

\section{Tagged Dataflow}
\label{tagged-dataflow}
The \emph{dataflow model of computation}~\cite{Davis78,DennisM74} was developed
more than forty years ago, as an alternative to the classical ``von-Neumann''
computing model. The key motivation was the creation of architectures and
programming languages that would exploit the massive parallelism that is
inherent in many applications. A dataflow program is essentially a directed
graph in which vertices correspond to processing elements and edges correspond
to channels. The data that need to be processed start ``flowing'' inside the
channels; when they reach a node they are being processed and the data produced
are fed to the output channels of the node. Since various parts of the dataflow
graph can be working concurrently, the parallel nature of the model should be
apparent. Moreover, this processing of data ``while in motion'' comes in sharp
contrast with the traditional ``von-Neumann'' model in which data wait passively
in memory until they are fetched by the central processing unit of the
(sequential) computer in order to be processed.

\subsection{Dataflow Graphs, Tags, and Tokens}
A key notion in our discussion is that of a \emph{dataflow graph}:
\begin{definition}\label{dataflow-graph}
A \emph{dataflow graph} (or \emph{dataflow network}) is a directed graph
$G=(V,E)$, where $V$ is the finite set of \emph{nodes} of the graph and $E$ is
the set of edges connecting elements of $V$. The set $V$ is partitioned into
disjoint subsets $V_I$ (\emph{input nodes}), $V_O$ (\emph{output nodes}) and
$V_{P}$ (\emph{processing nodes}), subject to the following restrictions:
\begin{itemize}
\item Every input node has no incoming edges and has one outgoing edge towards a
      processing node.

\item Every output node has no outgoing edges and has one incoming edge from a
      processing node.

\item Every processing node has incoming edges (at least one) from input nodes
      and/or from other processing nodes and outgoing edges (at least one) to
      output nodes and/or other processing nodes.
\end{itemize}
\end{definition}
Intuitively, input nodes provide the input data to a dataflow graph, processing
nodes perform the processing of data, and output nodes collect the
output data produced by the network.

In the initial dataflow model, channels were assumed to be unbounded FIFO
queues, i.e., the data were assumed to flow in a specific order inside the
channels. However, it soon became apparent that a model that would not impose
any particular temporal ordering of the data would be much more general. This
resulted in the so-called \emph{tagged dataflow}
model~\cite{WatsonG79,ArvindC83,mit-dataflow}. The basic idea behind tagged
dataflow is that data can flow inside a network accompanied by \emph{tags}
(i.e., labels). The tags can also carry essential information that can be used
in order to implement iterative and recursive algorithms.

Intuitively, edges of our dataflow networks carry tuples of the form
$\tok{t}{d}$ where $d$ is an element of a data domain $D$ and $t$ is an element
of a set of tags $T$. The set $T$ may be quite involved; in its simplest form it
can be a set of natural numbers, or in more demanding cases it can be the set of
lists of naturals numbers, etc. Pairs of the form $\tok{t}{d} \in T \times D$
are usually referred in the dataflow literature as \emph{tokens}.

\subsection{Tagged Dataflow Operations}
We now describe the operation of the nodes of a dataflow graph. We start with
the simplest case, namely input nodes. Such nodes carry a constant value which
they pass to their output. For example, an input node labeled with the natural
number $3$, passes in its output the value 3. Since we have adopted the
convention that edges carry tokens, the value 3 originating from an input node,
can be interpreted as a tuple $\tok{t}{3}$, for all possible tags $t$.

The operation of processing nodes (such as for example, nodes that perform
addition, multiplication, and so on), is straightforward
(see~\cite{mit-dataflow}): every such operator can \emph{fire} when in all of its
inputs there exist tokens that have exactly the same tag. When the operator
fires, it produces a token that has exactly the same tag as the input tokens
that it has ``consumed''. As an example, the operation of binary addition can be
described as follows:
\begin{flalign*}
  \sem{+}\ (\tok{t}{a}, \tok{t}{b})\ & =  \tok{t}{a+b}
\end{flalign*}
There exist operators that do not need data in all their input tokens in order
to produce output or that do not produce data in all their output tokens. We
will use the symbol $\ast$ to denote the absence of data. Historically, $\ast$
was named a ``hiaton'' in~\cite{lucid-book} where it is mentioned that ``a
hiaton can be thought as the `output' of a process which at a particular point
in time has no `genuine' data to send on''. In the TensorFlow terminology, a
hiaton is referred as a ``dead'' data item~\cite{tensorflow}. Using the concept
of hiaton, we can define the semantics of two common dataflow operators:
%
%
\begin{flalign*}
  \sem{\op{Switch}}\ (\tok{t}{\mathrm{true}}, \tok{t}{d})\ &
                                                = (\tok{t}{d},\tok{t}{\ast})\\
  \sem{\op{Switch}}\ (\tok{t}{\mathrm{false}}, \tok{t}{d})\ &
                             = (\tok{t}{\ast}, \tok{t}{d})\\
  \sem{\op{Merge}}\ (\tok{t}{d}, \tok{t}{\ast}) & = \tok{t}{d} \\
  \sem{\op{Merge}}\ (\tok{t}{\ast}, \tok{t}{d}) & = \tok{t}{d}
\end{flalign*}

We assume that all the operators we will be using, with the only exception of
\op{Merge}, have the following property: in order to fire, the tokens in their
inputs must contain data values that are not hiatons. If some of their data
inputs are hiatons then the hiatons are propagated to their outputs. The
propagation of hiatons, also referred as ``deadness propagation''
\cite{tensorflow}, is also useful in practice and especially in distributed
execution (see Section~\ref{dist-exec}).

\subsection{Tagged Iteration}
Iterative algorithms can be represented as cyclic dataflow graphs where the
output of one iteration becomes input of the next one. In order to increase
parallelism, we would like to allow the nodes of our dataflow graphs to process
concurrently data that belong to different iteration cycles. The tags can be
used to ensure the implementation of asynchronous iteration in an elegant and
effective way. The main idea is precisely defined in the following excerpt
from~\cite{manchester-dataflow}:
\begin{quote}
  {\em Each separate (loop) iteration reuses the same code but with different
  data. To avoid any confusion of operands from the different iterations, each
  data value is tagged with a unique identifier known as the iteration level
  that indicates its specific iteration.}
\end{quote}
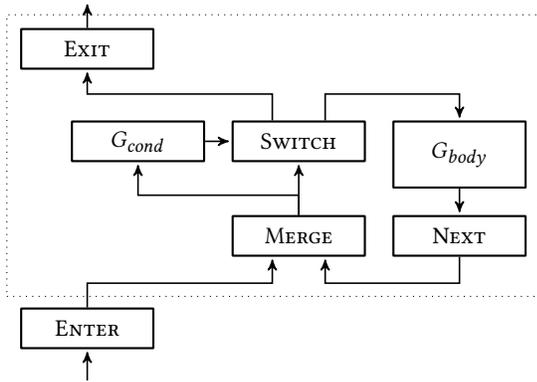
\begin{figure}
  \begin{tikzpicture}[>=latex]
  \tikzstyle{dataflow} = [draw, rounded corners=.2, thick,
           minimum height=1.5em, minimum width=5em, node distance=3.5em and 2.5em,
           ];
  \tikzstyle{to} =[->,>=stealth', shorten >=1pt, semithick];
  \tikzstyle{empty} = [draw=none, inner sep=.3cm],

  \node[empty] (out) {};
  \node[dataflow, below=1em of out] (exit) {\small \textsc{Exit}};
  \node[dataflow, below of=exit, xshift=8em](switch) {\small \textsc{Switch}};
  \node[dataflow, left=1em of switch] (cond) {\small $G_\mathit{cond}$};
  \node[dataflow, right=1em of switch, minimum height=2.5em, yshift=-0.5em] (body) {\small $G_\mathit{body}$};
  \node[dataflow, below= 2em of switch] (merge) {\small \textsc{Merge}};
  \node[dataflow, right=1em of merge] (next) {\small \textsc{Next}};
  \node[dataflow, below of=merge, xshift=-8em] (enter) {\small \textsc{Enter}};
  \node[empty, below of=enter] (in) {};

  \draw[to] (in)     -- (enter);
  \draw[to] (enter.north)  |- ($(merge.south)-(1em,1em)$) -- ($(merge.south)+(-1em,0)$);
  \draw[to] (merge)  -- (switch);
  \draw[to] (merge.north)  -- ++(0,0.75em) -| (cond);
  \draw[to] (cond)   -- (switch);
  \draw[to] ($(switch.north)+(-1em,0)$) -- ++(0,1em) -| (exit.south);
  \draw[to] ($(switch.north)+(1em,0)$) -| ++(0,+1em) -| (body);

  \draw[to] (body)   -- (next);
  \draw[to] (next.south) -- ++(0,-1em) -| ($(merge.south)+(1em,0)$);

  \draw[to] (exit)   -- (out);

  \begin{pgfonlayer}{background}
    \draw [join=round,black,dotted] ($(exit.north west) - (0.5em, -0.5em)$) rectangle ($(next.south east) + (0.5em, -1.5em)$);
  \end{pgfonlayer}

\end{tikzpicture}
  \caption{Dataflow implementation of a while-loop}
  \label{iteration}
\end{figure}
TensorFlow implements in a simple way the above idea. A tag in TensorFlow is a
list of natural numbers. The length of the list corresponds to the nesting of
the iteration. For example, if we have two nested while-loops, a list of length
2 signifies that we are inside the inner ``while''. Each item of the list
corresponds to the iteration counter inside a particular loop. So, for example,
a list of the form [2,3] means that we are inside two nested loops, the outer
loop has executed 3 times and the inner loop 2 times. In order to manage these
tags, TensorFlow supports the following operators:\footnote{We use the notation
$(n:t)$ to represent a list where $n$ is its first element (the \emph{head} of
the list) and $t$ is also a list (the \emph{tail} of the list $(n:t)$).}
\begin{flalign*}
  \sem{\op{Enter}}\ \tok{t}{d}  & = \tok{0:t}{d}  \\
  \sem{\op{Exit}}\ \tok{n:t}{d} & = \tok{t}{d}    \\
  \sem{\op{Next}}\ \tok{n:t}{d} & = \tok{(n+1):t}{d}
\end{flalign*}
Intuitively, the \op{Enter} operator is used when we enter a new while-loop. It
simply adds a new element in the list tag, representing in this way that the
level of nesting has been increased by 1. Symmetrically, \op{Exit} is used when
we exit a while-loop, and it obviously signals that the level of nesting has
just been decreased by 1. Finally, \op{Next} is used to increase the iteration
counter for the while-loop we are currently in. Figure~\ref{iteration} depicts a
dataflow graph that uses these operators in order to implement a simple
while-loop.

\subsection{Functions and Recursion}

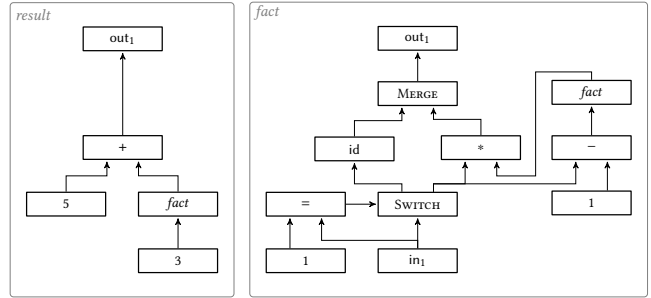
\begin{figure}
  \resizebox{0.48\textwidth}{!}{%
    \begin{tikzpicture}[>=latex]
  \tikzstyle{dataflow} = [draw, rounded corners=.2, thick,
           minimum height=1.5em, minimum width=5em, node distance=3.5em and 2.5em,
           ];
  \tikzstyle{empty} = [draw=none, inner sep=.3cm];
  \tikzstyle{to} =[->,>=stealth', shorten >=1pt, semithick];
  \tikzstyle{control} =[->,>=stealth', shorten >=1pt, dashed];
  \tikzstyle{func} =[draw, shape=rectangle, rounded corners=0.5ex, gray, text=gray,
      minimum width=2.5cm, text width=2.5cm, align=right, inner sep=10pt, inner ysep=15pt,label={[anchor=south west,yshift=-15pt,text=gray]north west:#1}];

  \node[dataflow] (out0) {\small $\mathsf{out}_1$};
  \node[dataflow, below of=out0] (merge-if) {\small \textsc{Merge}};
  \node[dataflow, below of=merge-if, xshift=-4em] (if-id) {\small $\mathsf{id}$};
  \node[dataflow, below of=if-id, xshift=4em] (switch) {\small \textsc{Switch}};
  \node[dataflow, above=2em of switch, xshift=+11em] (minus) {\small $-$};
  \node[dataflow, below of=minus] (const1-1) {\small $1$};
  \node[dataflow, above=2em of switch, xshift=+4em] (mult) {\small $*$};
  \node[dataflow, left= 2em of switch] (cond) {\small $=$};
  \node[dataflow, below=2em of cond] (const1) {\small $1$};
  \node[dataflow, below=2em of switch] (param) {\small $\mathsf{in}_1$};
  \node[dataflow, above of=minus] (fact) {\small $\mathit{fact}$};

  \draw[to] (cond) -- (switch);
  \draw[to] ($(const1.north) - (1em,0)$) -- ($(cond.south) - (1em,0)$);
  \draw[to] (param.north) -- ++ (0,0.5em) -| ($(cond.south) + (1em,0)$);
  \draw[to] (param.north) -- (switch);
  \draw[to] ($(switch.north) - (1em,0)$) -- ++ (0,0.5em) -| (if-id.south);
  \draw[to] (if-id.north) -- ++ (0,1em) -| ($(merge-if.south) - (1em,0)$);
  \draw[to] (mult.north) -- ++ (0,1em)  -| ($(merge-if.south) + (1em,0)$);

  \draw[to] ($(switch.north) + (1em,0)$) -- ++ (0,0.5em) -| ($(mult.south)-(1em,0)$);
  \draw[to] ($(switch.north) + (1em,0)$) -- ++ (0,0.5em) -| ($(minus.south)-(1em,0)$);
  \draw[to] ($(const1-1.north) + (1em,0)$) -- ($(minus.south)+(1em,0)$);

  \draw[to] (merge-if.north) -- ++(0,1em) -| (out0);

  \draw[to] (minus) -- (fact);
  \draw[to] (fact.north) -- ++(0,0.5em) -- ($(fact.north west)-(1em,-0.5em)$) |-
  ($(mult.south)+(1em,-1em)$) -- ($(mult.south)+(1em,0)$);

  \begin{pgfonlayer}{background}
    \node [func={$\mathit{fact}$}, fit=(const1) (out0) (fact)] (func_fact) {};
  \end{pgfonlayer}

  \node[dataflow, left=3em of const1] (const3) {\small $3$};
  \node[dataflow, above=2em of const3] (fact2) {\small $\mathit{fact}$};
  \node[dataflow, above=2em of fact2, xshift=-3.5em] (plus) {\small $+$};
  \node[dataflow, left=2em of fact2] (const5) {\small $5$};
  \node[dataflow, above=5.3em of plus] (out1)   {\small $\mathsf{out}_1$};

  \draw[to] (plus) -- (out1);
  \draw[to] (const3) -- (fact2);
  \draw[to] (const5.north) -- ++ (0,1em) -| ($(plus.south) - (1em,0)$);
  \draw[to] (fact2.north) -- ++ (0,1em) -| ($(plus.south) + (1em,0)$);

  \begin{pgfonlayer}{background}
    \node [func={$\mathit{result}$}, fit=(const5) (out1) (const3)] (func_result) {};
  \end{pgfonlayer}

\end{tikzpicture}}
  \caption{A simple program that computes the factorial expressed as a set of
           dataflow graphs}
  \label{fact-orig}
\end{figure}

The introduction of functions and recursion in the tagged dataflow framework is
more demanding than that of iteration. Supporting a set of (possibly recursive)
function definitions, implies that we have to use a \emph{set} of dataflow
graphs. As an example, consider the following simple recursive program computing
the factorial of a given number, consisting of two definitions. We use
Haskell-like functional notation:
\begin{verbatim}
   result  = fact(3)+5
   fact(n) = if (n==1) then n else n*fact(n-1)
\end{verbatim}
In order to represent this program as a dataflow graph, we need to create two
graphs, one for the variable {\tt result} and another for the function {\tt
fact}. These two graphs are depicted in Figure~\ref{fact-orig}.
In order to represent recursion, each separate graph is assigned a name ({\tt
result} and {\tt fact} in our example), which is the name of the function that
it implements. The dataflow graphs can then use nodes that contain these names,
the intuition being that these nodes correspond to recursive calls of the
corresponding functions. One can think of each such node as a ``black box'' that
hides inside it a nested graph of the function, which may itself hide inside it
another graph, and so on. This intuition seems to suggest that this view of
recursion requires a different graph structure than the one implied by
Definition~\ref{dataflow-graph}, namely a possibly infinite graph, or a
dynamically expanding one. If one wants to remain faithful to the
``fixed-graph'' approach of Definition~\ref{dataflow-graph}, a different
approach for handling recursion must be employed. Such an approach is introduced
in the next section.

\section{A Tagged Implementation of Recursion} \label{transformation}
In this section we demonstrate how first-order recursive functions can be
implemented in a purely tagged manner. Our construction builds on previous work
that was developed several years ago~\cite{WatsonG79,ArvindC83,mit-dataflow,
manchester-dataflow,yaghi,jfp1}, but has distinguishing characteristics from
each one of them. A detailed comparison of our approach with the aforementioned
ones, is given in Section~\ref{related-work}.

Assume we are given a set of dataflow graphs representing a program consisting
of (possibly) recursively defined functions. In this section we demonstrate how
this set of graphs can be transformed into a single {\em fixed} graph that can
be executed in a tagged way (namely, without the need of dynamic expansion). The
key idea is that whenever a function is called, we create a unique new tag that
characterizes this specific function invocation; as soon as we return from this
function call, we restore the tag to its previous state.

\begin{figure}
  \centering
  \resizebox{0.48\textwidth}{!}{%
    \begin{tikzpicture}[>=latex]
  \tikzstyle{dataflow} = [draw, rounded corners=.2, thick,
           minimum height=1.5em, minimum width=5em, node distance=3.5em and 2.5em,
           ];
  \tikzstyle{to} =[->,>=stealth', shorten >=1pt, semithick],
  \tikzstyle{control} =[->,>=stealth', shorten >=1pt, dashed],
  \tikzstyle{empty} = [draw=none, inner sep=.3cm],

  \node[empty] (out0) {};
  \node[dataflow, below=1em of out0] (ret0) {\small ${\textsc{Return}}_0$};
  \node[dataflow, below=1em of out0, xshift=8em] (ret1) {\small ${\textsc{Return}}_1$};
  \node[dataflow, below of=ret0, xshift=+4em] (merge-if) {\small \textsc{Merge}};
  \node[dataflow, below of=merge-if, xshift=-4em] (if-id) {\small $\mathsf{id}$};
  \node[dataflow, below of=if-id, xshift=4em] (switch) {\small \textsc{Switch}};
  \node[dataflow, below of=merge-if, xshift=+11em] (minus) {\small $-$};
  \node[dataflow, below of=minus] (const1-1) {\small $1$};
  \node[dataflow, below of=merge-if, xshift=+4em] (mult) {\small $*$};
  \node[dataflow, left= 2em of switch] (cond) {\small $=$};
  \node[dataflow, below=2em of cond] (const1) {\small $1$};
  \node[dataflow, below=2em of switch] (merge) {\small \textsc{Merge}};
  \node[dataflow, above of=minus] (call1) {\small $\textsc{Call}_1$};

  \draw[to] (cond) -- (switch);
  \draw[to] ($(const1.north) - (1em,0)$) -- ($(cond.south) - (1em,0)$);
  \draw[to] (merge.north) -- ++ (0,0.5em) -| ($(cond.south) + (1em,0)$);
  \draw[to] (merge) -- (switch);
  \draw[to] ($(switch.north) - (1em,0)$) -- ++ (0,0.5em) -| (if-id.south);
  \draw[to] (if-id.north) -- ++ (0,1em) -| ($(merge-if.south) - (1em,0)$);
  \draw[to] (mult.north) -- ++ (0,1em)  -| ($(merge-if.south) + (1em,0)$);

  \draw[to] ($(switch.north) + (1em,0)$) -- ++ (0,0.5em) -| ($(mult.south)-(1em,0)$);
  \draw[to] ($(switch.north) + (1em,0)$) -- ++ (0,0.5em) -| ($(minus.south)-(1em,0)$);
  \draw[to] ($(const1-1.north) + (1em,0)$) -- ($(minus.south)+(1em,0)$);

  \draw[to] (merge-if.north) -- ++(0,1em) -| (ret0);
  \draw[to] (merge-if.north) -- ++(0,1em) -|(ret1);

  \draw[to] (ret1.north) -- ++(0,0.5em) -- ($(ret1.north east)+(1em,0.5em)$) -- ($(mult.south east)+(1em,-1em)$) -| ($(mult.south)+(1em,0)$);

  \draw[to] (minus) -- (call1);
  \draw[to] (call1.north) -- ++(0,0.5em) -- ($(call1.north east)+(1em,0.5em)$) |-
  ($(merge.south)+(1em,-1em)$) -- ($(merge.south)+(1em,0)$);





  \node[dataflow, left=3em of const1] (const3) {\small $3$};
  \node[dataflow, above=2em of const3] (call0) {\small $\textsc{Call}_0$};
  \node[dataflow, above=2em of call0, xshift=-3.5em] (plus) {\small $+$};
  \node[dataflow, left=2em of call0] (const5) {\small $5$};
  \node[dataflow, above=5.3em of plus] (out1)   {\small $\mathsf{out}_1$};

  \draw[to] (plus) -- (out1);
  \draw[to] (const3) -- (call0);
  \draw[to] (const5.north) -- ++ (0,1em) -| ($(plus.south) - (1em,0)$);
  \draw[to] (call0.north) -- ++ (0,0.5em) -| ($(call0.north east)+(1em,0.5em)$) |- ($(merge.south)+(-1em,-1em)$) -- ($(merge.south)+(-1em, 0)$);
  \draw[to] (ret0.north) -- ++ (0,0.5em) -| ($(call0.north east)+(1em,1em)$) -| ($(plus.south)+(+1em,0)$);

  \draw[control] (call0) edge[bend left=-20] (ret0);
  \draw[control] (call1) edge[bend left=20] (ret1);
  \draw[control] (merge) -- (const1);
  \draw[control] (merge) edge[bend right=20] (const1-1.south);

  \begin{pgfonlayer}{background}
  \end{pgfonlayer}

\end{tikzpicture}}
  \caption{The transformed graph of $\mathit{fact}$}
  \label{fact-trans}
\end{figure}
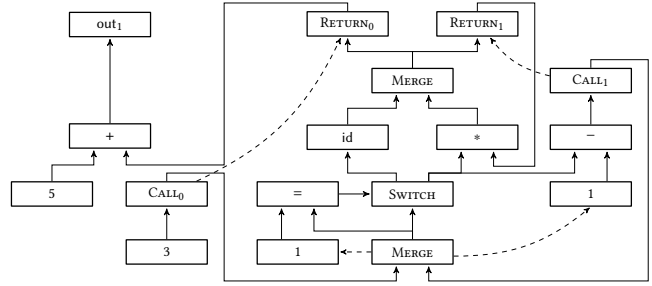

Before we present the algorithm in a formal way, we motivate it through a simple
example. Consider the two graphs of the factorial example depicted in
Figure~\ref{fact-orig}. In the following, we present each step of the algorithm,
together with the intuition behind each step:
\begin{itemize}
\item There are two nodes labeled with {\tt fact} in Figure~\ref{fact-orig}. We
      replace one of them with a node labeled $\op{Call}_0$ and the other one
      with a node labeled $\op{Call}_1$. It is not important which node receives
      which operator, as long as each different node receives a \op{Call}
      operator with a different subscript. Intuitively, each $\op{Call}_i$
      operator creates a unique new tag that characterizes the specific function
      call that it has replaced.

\item The input node of the {\tt fact} graph is replaced by a \op{Merge}
      node. The output edges of $\op{Call}_0$ and $\op{Call}_1$ become the two
      inputs of the \op{Merge} node. Intuitively, this \op{Merge} node gathers
      together all the different ``entries'' to the body of the function {\tt
      fact} (avoiding in this way to have different ``copies'' of the body of
      {\tt fact}).

\item The output node of the {\tt fact} graph is connected with two new
      nodes labeled as $\op{Return}_0$ and $\op{Return}_1$. The output edge of
      each $\op{Return}_i$ is connected to the node where the output of the
      corresponding {\tt fact} node was directed in the initial graphs.
      Intuitively, each $\op{Return}_i$ operator restores the tag to the state
      it had just before the corresponding $\op{Call}_i$.

\item We add an extra output edge to each $\op{Call}_i$ connecting it with its
      corresponding $\op{Return}_i$ node. Intuitively, this edge expresses the
      requirement that in order for a $\op{Return}_i$ node to fire, a
      $\op{Call}_i$ must have fired at some previous time point\footnote{The
      necessity of adding these edges will be explained in
      Section~\ref{deadness-propagation}.}. Notice that this is a
      ``synchronizing'' edge, ie., the data value that it carries is irrelevant.
      Such edges are termed {\em control edges} in the TensorFlow literature and
      are usually denoted with dashed lines.
\end{itemize}
The transformation described above, produces a single dataflow graph depicted in
Figure~\ref{fact-trans}. It remains to specify the intuitive and formal meaning
of the $\op{Call}_i$ and $\op{Return}_i$ operators. As we have already
discussed, the underlying idea behind the tagged implementation of recursion, is
to represent with a unique tag the data that belong to a specific invocation of
a function. In this way we avoid mixing data that belong to different
invocations. When the operator $\op{Call}_i$ takes as input a token
$\tok{t}{d}$, it changes the tag $t$ into a new one that represents the function
call that is being invoked. This is being done by prefixing $t$ with $i$,
namely:
\begin{align*}
  \sem{\op{Call}_i}\ \tok{t}{d} & =  \tok{i:t}{d}
\end{align*}
Intuitively, the list $i:t$ identifies the position in the recursion tree where
execution currently is. When the function returns, we must restore the tag to
the state it had before the function was called. This is performed by the
operator $\op{Return}_i$ which performs the converse operation from that of
$\op{Call}_i$:
\begin{align*}
  \sem{\op{Return}_i}\ \tok{i:t}{d} & = \tok{t}{d}
\end{align*}
Using the above semantic equations, one can easily ``run'' the resulting graph
depicted in Figure~\ref{fact-trans} and verify that it computes the desired
output.

The transformation described for the {\tt fact} program, can be easily
generalized to programs with many different functions. Assume we are given a set
$S$ of dataflow graphs representing a set of (possibly recursively) defined
functions. We assume that one of these function definitions is for a variable
{\tt result} (intuitively, the output of the program). The description is
slightly more complicated because we now consider functions that may have more
than one formal parameters. The transformation proceeds as follows:
\begin{itemize}

\item Let $f$ be a function defined in $S$ having $m$ formal parameters. Assume
      there are $n$ nodes in $S$ labeled with the name of $f$ (each such node
      corresponds to a different call to $f$). Number these nodes starting from
      0 up to $n-1$ (it is not important which node of $f$ receives which
      number, as long as each different node receives a different number).

\item Replace the $i$'th node of $f$ with $m$ identical nodes all labeled with
      $\op{Call}_i$, each one of them corresponding to one of the $m$ different
      formal parameters of $f$. Each one of these $\op{Call}_i$ nodes has only
      one input, namely the input to the original $f$ node that corresponded to
      the specific formal parameter of $f$.

\item In the initial graph of $f$, there exist $m$ input nodes $\mathsf{in}_0,
      \ldots, \mathsf{in}_{m-1}$, each one corresponding to a different formal
      parameter of $f$. Replace each such input node by a \op{Merge} node.
      Direct the output edge of each $\op{Call}_i$ node corresponding to a
      specific formal parameter of $f$, to the input of the corresponding
      \op{Merge} node.

\item The output of $f$'s graph is connected with $n$ new nodes
      labeled $\op{Return}_0, \ldots, \op{Return}_{n-1}$. The output edge of
      each $\op{Return}_i$ is connected to the node where the output of the
      $i$'th node of $f$ was connected in the initial version of the graph.

\item We add an extra output edge to each $\op{Call}_i$ connecting it with its
      corresponding $\op{Return}_i$ node.

\end{itemize}
The above algorithm transforms the given set $S$ into a new graph; the output of
this graph is the output node of the graph corresponding to the variable {\tt
result}.


\section{Automatic Differentiation} \label{autodiff}

Neural networks are typically trained using the backpropagation
algorithm~\cite{backprop}. This involves the minimization of a loss function
that naturally depends on gradient-based computations. Dataflow systems that
focus on deep learning applications typically implement the backpropagation
algorithm via a systematic way called reverse-mode automatic
differentiation~\cite{autodiff-survey}. Roughly speaking, automatic
differentiation derives automatically the operations that need to take place in
order to compute the gradient of the function that is encoded as a dataflow
graph. In this section we describe how we support automatic differentiation in
recursive dataflow graphs that have been transformed using the tagging approach.

The automatic differentiation algorithm uses the chain rule to connect newly
introduced gradient operators (i.e., operators that compute the gradients) to
the graph. More specifically, for a simple tagless dataflow graph, for each
operator \((y_1, \ldots, y_m) = \mathsf{op_i}(x_1, \ldots, x_n) \) the algorithm
adds a corresponding gradient operator $\mathsf{\grad{op_i}}$ that computes the
gradients $dx_1, \ldots, dx_n$ of the inputs with respect to the gradients
$dy_1, \ldots, dy_m$ of the outputs of $\mathsf{op_i}$. Note that in general
$\mathsf{\grad{op_i}}$ also requires the original inputs $x_1, \ldots, x_n$ of
$\mathsf{op_i}$ in order to operate. Hence, $\mathsf{\grad{op_i}}$ is an
operator that accepts $n + m$ inputs and produces $n$ outputs. This process
results into a dataflow graph that is naturally activated in two phases: the
forward and backward phase. The forward phase involves the activation of the
original operators of the graph that compute the original outputs (also called
forward values) of the dataflow. When all the input values of the last operator
are available the backward phase can start to compute the gradient values. Note
that, since the gradient operators need the input values of the original
operator, the values produced in the forward phase are usually retained in
memory during the whole execution to avoid unnecessary recomputations.

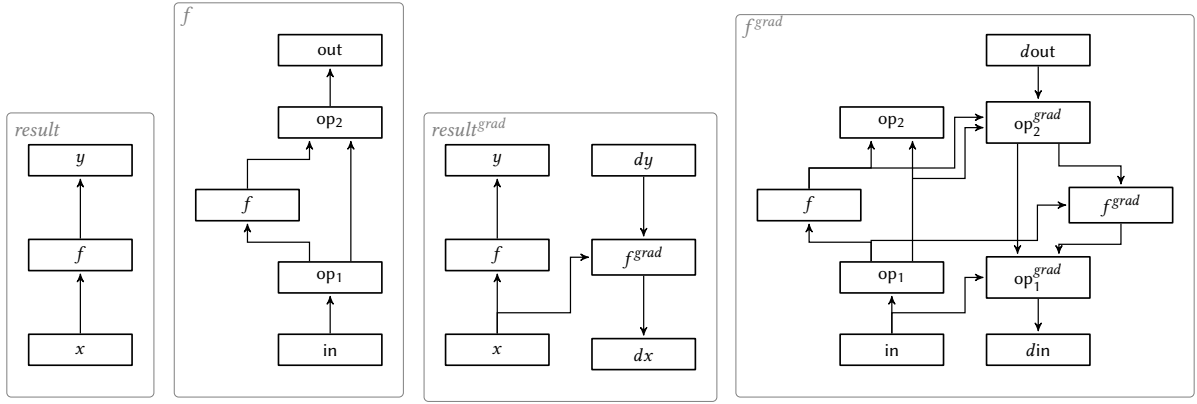
\begin{figure*}
  \resizebox{0.9\textwidth}{!}{
    \begin{tikzpicture}[>=latex]
  \tikzstyle{dataflow} = [draw, rounded corners=.2, thick,
           minimum height=1.5em, minimum width=5em, node distance=3.5em and 2.5em,
           ];
  \tikzstyle{empty} = [draw=none, inner sep=.3cm];
  \tikzstyle{to} =[->,>=stealth', shorten >=1pt, semithick];
  \tikzstyle{control} =[->,>=stealth', shorten >=1pt, dashed];
  \tikzstyle{func} =[draw, shape=rectangle, rounded corners=0.5ex, gray, text=gray,
      minimum width=2.5cm, minimum height=2.5cm, text width=2.5cm, align=left,
      inner sep=10pt, inner ysep=15pt,label={[anchor=south west,yshift=-15pt,text=gray]north west:#1}];


  \node[dataflow] (in) {\small $x$};
  \node[dataflow, above=3em of in] (f) {\small $f$};
  \node[dataflow, above=3em of f] (y) {\small $y$};
  \draw[to] (in) -- (f);
  \draw[to] (f)  -- (y);

  \begin{pgfonlayer}{background}
    \node [func={$\mathit{result}$}, fit=(y) (in)] (func_result) {};
  \end{pgfonlayer}

  \node[dataflow, right=7em of in] (fin) {\small $\mathsf{in}$};
  \node[dataflow, above of = fin] (fop1) {\small $\mathsf{op_1}$};
  \node[dataflow, above of = fop1, xshift=-4em] (ff) {\small $f$};
  \node[dataflow, above of = fop1, yshift=4em] (fop2) {\small $\mathsf{op_2}$};
  \node[dataflow, above of = fop2] (fout) {\small $\mathsf{out}$};

  \draw[to] (fin) -- (fop1);
  \draw[to] ($(fop1.north) - (1em,0)$) |-  ($(ff.south) - (0,1em)$) -- (ff);
  \draw[to] (ff)  |- ($(fop2.south) - (1em,1em)$) -- ($(fop2.south) - (1em,0)$);
  \draw[to] (fop2)  -- (fout);
  \draw[to] ($(fop1.north) + (1em,0)$) -- ($(fop2.south) + (1em,0)$);

  \begin{pgfonlayer}{background}
    \node [func={$f$}, fit=(fout) (ff) (fin)] (func_f) {};
  \end{pgfonlayer}

  \node[dataflow, right=3em of fin] (gin) {\small $x$};
  \node[dataflow, above=3em of gin] (gf) {\small $f$};
  \node[dataflow, above=3em of gf] (gy) {\small $y$};
  \node[dataflow, right=2em of gy] (gdy) {\small $dy$};
  \node[dataflow, below=3em of gdy] (gdf) {\small $\mathit{f^\mathit{grad}}$};
  \node[dataflow, below=3em of gdf] (gdin) {\small $dx$};
  
  \draw[to] (gin) -- (gf);
  \draw[to] (gf)  -- (gy);
  \draw[to] (gin.north)  -- ($(gin.north) + (0,1em)$) -| ($(gdf.west) -
  (1em,0)$) -- (gdf.west);
  \draw[to] (gdy) -- (gdf);
  \draw[to] (gdf) -- (gdin);

  \begin{pgfonlayer}{background}
    \node [func={$\mathit{result}^\mathit{grad}$}, fit=(gy) (gdin)] (func_gresult) {};
  \end{pgfonlayer}

  \node[dataflow, right=14em of gin] (fgin) {\small $\mathsf{in}$};
  \node[dataflow, above of = fgin] (fgop1) {\small $\mathsf{op_1}$};
  \node[dataflow, above of = fgop1, xshift=-4em] (fgf) {\small $f$};
  \node[dataflow, above of = fgop1, yshift=4em] (fgop2) {\small $\mathsf{op_2}$};
  \node[empty, above of = fgop2] (fgout) {};
  \node[dataflow, right= 2em of fgop2] (dfgop2) {\small $\mathsf{op_2^\mathit{grad}}$};
  \node[dataflow, above of = dfgop2] (dfgout) {\small $d\mathsf{out}$};
  \node[dataflow, below of = dfgop2, yshift=-4em] (dfgop) {\small $\mathsf{op_1^\mathit{grad}}$};
  \node[dataflow, above of = dfgop, xshift=4em] (dfg) {\small $f^\mathit{grad}$};
  \node[dataflow, below of = dfgop] (dfgin) {\small $d\mathsf{in}$};

  \draw[to] (fgin) -- (fgop1);
  \draw[to] ($(fgop1.north) - (1em,0)$) |- ($(fgf.south) - (0,1em)$) -- (fgf);
  \draw[to] (fgf) -- ($(fgf.north) + (0,1em)$)  -| ($(fgop2.south) - (1em,0)$);
  \draw[to] ($(fgop1.north) + (1em,0)$) -- ($(fgop2.south) + (1em, 0)$);
  \draw[to] (dfgout) -- (dfgop2);
  \draw[to] ($(dfgop2.south) + (1em,0)$) |- ($(dfg.north) + (0,1em)$) -- (dfg);
  \draw[to] (dfg) -- ($(dfg.south) - (0,1em)$) -| ($(dfgop.north) + (1em,0)$);
  \draw[to] ($(dfgop2.south) - (1em,0)$) -- ($(dfgop.north) - (1em,0)$);
  \draw[to] (dfgop) -- (dfgin);
  \draw[to] (fgin.north)  -- ($(fgin.north) + (0,1em)$) -| ($(dfgop.west) -
  (1em,0)$) -- (dfgop.west);
  \draw[to] ($(fgop1.north) - (1em,0)$) -- ($(fgop1.north) + (-1em,1em)$) -|
  ($(dfg.west)-(1.5em,0)$) -- (dfg.west);
  \draw[to] (fgf) -- ($(fgf.north) + (0,1em)$) -| ($(dfgop2.west) - (1.5em,
  -0.25em)$) -- ($(dfgop2.west) - (0,-0.25em)$);

\draw[to] ($(fgop1.north) + (1em,4em)$) -| ($(dfgop2.west) - (1em,
  +0.25em)$) -- ($(dfgop2.west) - (0,+0.25em)$);

  \begin{pgfonlayer}{background}
    \node [func={$f^\mathit{grad}$}, fit= (fgf) (fgout) (dfgin) (dfg)] (func_dfg) {};
  \end{pgfonlayer}

\end{tikzpicture}
  }
\caption{The set of dataflow graphs that compute the gradient of {\tt result}.}
\label{autodiff-functions}
\end{figure*}
%
In the simple case we have discussed so far the graph consists of primitive
operators only. The corresponding gradient operators are predefined and the
mapping between the operator and its gradient counterpart is typically
maintained in a catalogue for the purpose of the automatic differentiation.
However, if we want to consider dataflow graphs that contain user-defined
functions as operators, we also need to consider how to generate the appropriate
gradient operators of these functions. Recall that, in this setting, each
function is defined as a separate dataflow graph and as such can be
automatically differentiated. Therefore, the straightforward approach is to
derive $\grad{f}$ by applying the automatic differentiation algorithm to the
dataflow graph of $f$. This will produce a new graph that effectively computes
the gradient of $f$. The derived graph can then be added as an ordinary function
definition $\grad{f}$ and used as the gradient operator of $f$.
Figure~\ref{autodiff-functions} depicts a set of dataflow graphs consisting of
the main graph named $\mathit{result}$ that uses a recursive function $f$. The
corresponding $\grad{f}$ is itself recursive, and is produced by automatically
differentiating $f$. Also note that, as expected, the gradient computation of
$\mathit{result}$ uses both $f$ and $\grad{f}$.

It is easy to see that the resulting set of dataflow graphs in
Figure~\ref{autodiff-functions} can be transformed to a single static dataflow
graph using the transformation algorithm introduced in
Section~\ref{transformation}. However, this approach would be highly
inefficient: the invocation of $\grad{f}$ triggers the evaluation of the forward
values of $f$ that have also been computed some time in the past by the
invocation of $f$. In the specific example, the operators $\mathsf{op_1}$ and
$\mathsf{op_2}$ will execute in both invocations of $f$ and $\grad{f}$ in
$\grad{\mathit{result}}$. The main reason is that, using this naive process for
automatic differentiation, we treat the gradient computation as a separate
function and therefore we lose the relationship between the original function
invocation and its corresponding gradient computation.

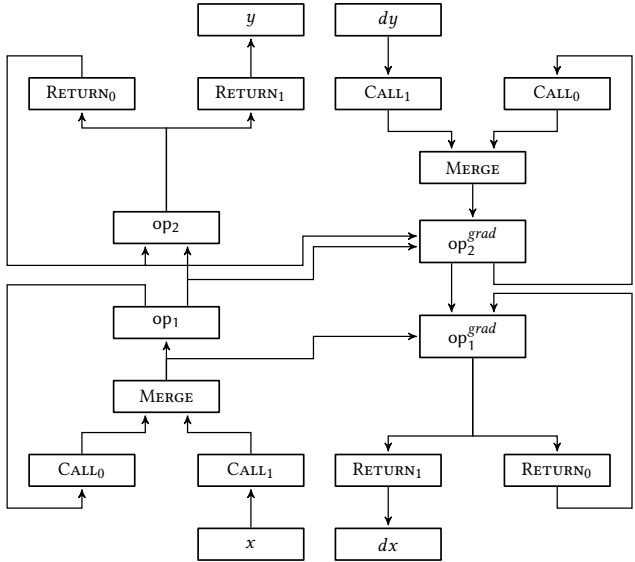
\begin{figure}
  \resizebox{0.48\textwidth}{!}{
    \begin{tikzpicture}[>=latex]
  \tikzstyle{dataflow} = [draw, rounded corners=.2, thick,
           minimum height=1.5em, minimum width=5em, node distance=3.5em and 2.5em,
           ];
  \tikzstyle{empty} = [draw=none, inner sep=.3cm];
  \tikzstyle{to} =[->,>=stealth', shorten >=1pt, semithick];
  \tikzstyle{control} =[->,>=stealth', shorten >=1pt, dashed];
  \tikzstyle{func} =[draw, shape=rectangle, rounded corners=0.5ex, gray, text=gray,
      minimum width=2.5cm, minimum height=2.5cm, text width=2.5cm, align=left,
      inner sep=10pt, inner ysep=15pt,label={[anchor=south west,yshift=-15pt,text=gray]north west:#1}];

  \node[dataflow] (merge_orig) {\small \op{Merge}};
  \node[dataflow, below of=merge_orig, xshift=-4em] (call1_orig) {\small $\op{Call}_0$ };
  \node[dataflow, below of=merge_orig, xshift=4em] (call2_orig) {\small $\op{Call}_1$ };

  \node[dataflow, above of=merge_orig] (op1_orig) {\small $\mathsf{op_1}$};
  \node[dataflow, above of=op1_orig, yshift=1em] (op2_orig) {\small $\mathsf{op_2}$};
  \node[empty, above of=op2_orig] (split_orig) {};
  \node[dataflow, above of=split_orig, xshift=-4em] (ret1_orig) {\small $\op{Return}_0$};
  \node[dataflow, above of=split_orig, xshift=4em] (ret2_orig) {\small $\op{Return}_1$};
  
  \draw[to] (call1_orig.north) -- ($(call1_orig.north) + (0,1em)$) -| ($(merge_orig.south) - (1em,0)$);
  \draw[to] (call2_orig.north) -- ($(call2_orig.north) + (0,1em)$) -|
  ($(merge_orig.south) + (1em,0)$);
  \draw[to] (merge_orig) -- (op1_orig);
  \draw[to] (op2_orig.north) -- (split_orig.north) |- ($(split_orig.north) - (1em,-1em)$) -| (ret1_orig);
  \draw[to] (op2_orig.north) -- (split_orig.north) |- ($(split_orig.north) + (1em,1em)$) -| (ret2_orig);

  \draw[to] (ret1_orig.north) -- ($(ret1_orig.north) + (0,1em)$) -|
  ($(ret1_orig.north west) - (1em,1em)$) |-
  ($(op2_orig.south) - (1em,1em)$) -- ($(op2_orig.south) - (1em,0)$);

  \draw[to] ($(op1_orig.north) - (1em,0)$) -- ($(op1_orig.north) + (-1em,1em)$) -| 
  ($(call1_orig.south west) - (1em,1em)$) -| (call1_orig.south);

  \draw[to] ($(op1_orig.north) + (1em,0)$) -- ($(op2_orig.south) + (1em,0)$);

  \node[dataflow, right of=split_orig, xshift=11em] (merge_grad) {\small
  \op{Merge}};
  \node[dataflow, above of=merge_grad, xshift=4em] (call1_grad) {\small $\op{Call}_0$};
  \node[dataflow, above of=merge_grad, xshift=-4em]  (call2_grad) {\small $\op{Call}_1$};
  \node[dataflow, below of=merge_grad] (op2_grad) {\small
  $\mathsf{op_2^\mathit{grad}}$};
  \node[dataflow, below of=op2_grad, yshift=-1em] (op1_grad) {\small $\mathsf{op_1^\mathit{grad}}$};
  \node[empty, below of=op1_grad] (split_grad) {};
  \node[dataflow, below of=split_grad, xshift=4em] (ret1_grad) {\small $\op{Return}_0$ };
  \node[dataflow, below of=split_grad, xshift=-4em] (ret2_grad) {\small $\op{Return}_1$ };
  
  \draw[to] (call1_grad.south) -- ($(call1_grad.south) - (0,1em)$) -| ($(merge_grad.north) + (1em,0)$);
  \draw[to] (call2_grad.south) -- ($(call2_grad.south) - (0,1em)$) -|
  ($(merge_grad.north) - (1em,0)$);
  \draw[to] (merge_grad) -- (op2_grad);
  \draw[to] (op1_grad) -- (split_grad.south) |- ($(split_grad.south) - (1em,1em)$) -| (ret2_grad);
  \draw[to] (op1_grad) -- (split_grad.south) |- ($(split_grad.south) + (1em,-1em)$) -| (ret1_grad);

  \draw[to] (ret1_grad.south) -- ($(ret1_grad.south) - (0,1em)$) -|
  ($(ret1_grad.south east) + (1em,1em)$) |-
  ($(op1_grad.north) + (1em,1em)$) -- ($(op1_grad.north) + (1em,0)$);

  \draw[to] ($(op2_grad.south) + (1em,0)$) -- ($(op2_grad.south) - (-1em,1em)$) -| 
  ($(call1_grad.north east) + (1em,1em)$) -| (call1_grad.north);

  \draw[to] ($(op2_grad.south) - (1em,0)$) -- ($(op1_grad.north) - (1em,0)$);

  \draw[to] (merge_orig.north) -- ($(merge_orig.north) + (0,1em)$) -|
  ($(op1_grad.west) - (5em,0)$) -- (op1_grad.west);
  \draw[to] ($(op1_orig.north) + (1em,1.25em)$) -|
  ($(op2_grad.west) - (5em,0.25em)$) -- ($(op2_grad.west) - (0,0.25em)$);
  \draw[to] ($(op2_orig.south) - (1em,1em)$) -|
  ($(op2_grad.west) - (5.5em,-0.25em)$) -- ($(op2_grad.west) - (0,-0.25em)$);

  \node[dataflow, below of = call2_orig] (result_in) {\small $x$ };
  \draw[to] (result_in) -- (call2_orig);
  \node[dataflow, below of = ret2_grad] (result_din) {\small $dx$ };
  \draw[to] (ret2_grad) -- (result_din);
  \node[dataflow, above of = ret2_orig] (result_out) {\small $y$ };
  \draw[to] (ret2_orig) -- (result_out);
  \node[dataflow, above of = call2_grad] (result_dout) {\small $dy$ };
  \draw[to] (result_dout) -- (call2_grad);
\end{tikzpicture}
  }
\caption{The transformed dataflow graph that computes the gradient of {\tt result}.}
\label{autodiff-transf}
\end{figure}
In order to remedy this inefficiency we have devised a more sophisticated way of
supporting automatic differentiation. The key idea is to identify each function
invocation $f$ and its counterpart invocation $\grad{f}$ in the graph and
replace them simultaneously (i.e., assign to them the same label $i$ for
$\op{Call}_i$-$\op{Return}_i$) with a single dataflow graph that computes both
the forward values and the gradients. This subgraph will have $n+m$ inputs,
corresponding to $n$ inputs of $f$ and $m$ gradient inputs of $\grad{f}$, and
$m+n$ outputs, corresponding to $m$ outputs of $f$ and $n$ gradient outputs of
$\grad{f}$. For example, the transformed graph of
Figure~\ref{autodiff-functions} is shown in Figure~\ref{autodiff-transf}. There
are two distinguished labels, one corresponding to the invocations in
$\grad{\mathit{result}}$ and the other to the invocations in $\grad{f}$.

An interesting observation is that, using this approach, the forward values
produced by the operators inside $f$ are communicated to their gradient
counterparts which naturally reside in $\grad{f}$. The use of the same tag for
both the original operators and the gradient operators, will ensure that the
correct forward values will be matched with the correct invocations of the
gradient operators.

\section{Implementation in TensorFlow} \label{impl}

In this section we discuss the implementation issues that arise in the tagged
approach to recursion introduced in the previous sections. As it turns out, the
existing infrastructure of TensorFlow requires relatively small changes in order
to incorporate the new ideas. In particular, it suffices to extend the
computational model of TensorFlow to accommodate the execution of dataflow
graphs that contain $\op{Call}_i$ and $\op{Return}_i$ operators. In the
following, we discuss the implementation details that arise in both the
single-machine and the distributed execution of these extended dataflow graphs.

\subsection{Local Execution} \label{local-exec}

TensorFlow executes a tagless dataflow graph in a simple way: it places in a
``ready'' queue those nodes that have all their inputs available, and the
remaining nodes in a ``pending'' queue. It starts by executing (possibly in
parallel) the nodes in the ``ready'' queue. The nodes from the ``pending'' queue
are moved to the ``ready'' queue when all their inputs become available. The
overall computation terminates when both queues become empty.

For dataflow graphs that use tags, TensorFlow allows multiple copies of a single
node to appear in the queues, each copy corresponding to a different tag (that
the inputs of the node must have in order for the node to fire). Execution
proceeds as in the case of tagless graphs. Tags are implemented in TensorFlow
with the help of \emph{frames}. A frame conceptually captures a specific scope
of execution (e.g., an iteration of a while-body). Frames are organized in
stacks, that is, each frame has a reference to the previous scope. Tags, are
essentially pairs of an iteration number and a pointer to that global stack. The
\op{Enter} node allocates a new frame initializing its value to 0; the \op{Exit}
node deallocates a frame\footnote{Actually, the operations performed by
\op{Enter} and \op{Exit} are more complicated than this, but for the purposes of
our discussion, a more precise description is unnecessary.}; the \op{Next} node
increases the value of the current frame by 1.

In our implementation we reuse the frame stack that already corresponds to the
static nesting scopes of while-loops to also incorporate the (dynamic) scope of
functions. The new primitive operators that we have added, namely $\op{Call}_i$
and $\op{Return}_i$, actually extend the already available operations \op{Enter}
and \op{Exit} of TensorFlow. Therefore, the required changes in the runtime of
TensorFlow, are small. More specifically, we extend frames to include a variable
in which the label $i$ of $\op{Call}_i$ is stored. Every time a $\op{Call}_i$
node is executed, a new frame is allocated and this local variable of the frame
is set to the value $i$. Similarly, when $\op{Return}_i$ is executed, the local
variable of the current frame is first checked; if it is equal to $i$, then the
frame is deallocated from the stack, otherwise no output is produced.

It is easy to see why the scopes of the functions and the scopes of the
while-loops can be fused into a single stack, without adding any extra tagging
machinery. Imagine, for example, a program that has a while-loop inside of which
a function is called. When the function is called, a new frame is allocated and
the dataflow subgraph corresponding to this function call is executed with
respect to this frame. When the execution of the function returns, the stack
frame is deallocated, and the execution of the dataflow graph corresponding to
the while-loop continues with respect to the tag that existed before the
function was invoked. In this way, recursive functions and iteration are
supported by the same simple tagging mechanism.

\subsection{Distributed Execution} \label{dist-exec}

Distributed execution can be achieved by partitioning the dataflow graph into
subgraphs and assigning them to different machines. Each machine then executes
the subgraph locally, using the same process described in
Section~\ref{local-exec}. The distributed computation completes when all the
machines finish.

TensorFlow imposes no restriction to the partitioning of the graph and as a
result parts of the same function may reside on different machines. In order not
to limit parallelism, TensorFlow employs a design that has no centralized
coordination of the different machines. Instead, the subgraphs that reside on
the different machines are extended with \op{Send} and \op{Recv} nodes that
enable direct communication between the machines. In order to support recursion,
we have extended this framework as described below.

\subsubsection{Deadness Propagation} \label{deadness-propagation}

When all the results of a distributed computation become available, one has to
ensure that all the machines that have participated in the computation, will
terminate. As an example where such a need arises, consider the case of the
subgraph of a conditional branch executing in a separate machine waiting for
input via a \op{Recv} node. If, at runtime, this branch is not selected, then
the machine will wait indefinitely and the execution will not terminate without
some intervention. This issue is tackled in TensorFlow by sending a ``dead''
token to explicitly signify the absence of data.

In our extended framework, we must ensure the correct deadness propagation in
the presence of the two new operators $\op{Call}_i$ and $\op{Return}_i$. In
particular, we ensure that the dead token received in a $\op{Call}_i$ will be
propagated to its corresponding $\op{Return}_i$, by adding a control edge from
the former node to the latter. In this way every $\op{Return}_i$ will receive a
``dead'' token, which it will also propagate to its successor nodes. This
process ensures the termination of all processes that are engaged in a
distributed computation.

\subsubsection{Tag Tracking} \label{tag-tracking}

Tag tracking is an issue that arises from the fact that nodes that operate on
the same tagged data can be distributed in several machines and connect directly
through a \op{Send}-\op{Recv} pair. Recall that in a local execution setting a
node is scheduled accompanied with a specific tag, that is the tag that the data
should match. An immediate consequence is that \op{Recv} nodes should also be
scheduled with respect to a certain tag that may have originally been generated
in a different machine (that is, the tag-generating nodes reside in a different
machine). This observation makes it apparent that a mechanism should be
established in order to track among the relevant machines which tags have been
generated. TensorFlow tackles this issue by creating a ``state machine'' in
every relevant machine that is essentially the backbone of the graph that
includes the tag-generating nodes (e.g. \op{Enter}, \op{Next}). This apparatus
simulates the decisions that occur distributively in the graph (that is, which
branch in \op{Switch} nodes is active) and replays the tag creation process with
dummy input data to all the machines that do not originally have those nodes.

In the case of the tagged dataflow graphs with recursion, the form of the
``state machine'' is slightly more complex but similar. In order to extract the
correct form of the state machines we do the following:
\begin{itemize}

\item We traverse the graph and capture the occurrences of the tag-generating
      nodes. As tag-generating we consider $\op{Call}_i$, \op{Enter} and
      \op{Next}.

\item We also capture the occurrences of the control-flow decision nodes, namely
      the \op{Switch} and \op{Merge} nodes.

\item We generate a new graph using copies of the captured nodes. The edges of
      the graph are inferred by the original graph if we suppress the remaining
      nodes.

\item We introduce a control edge between a tag-generating node $n$ and the
      \op{Recv} nodes that connect nodes after node $n$ and before the next
      tag-generating node. Intuitively these \op{Recv} nodes should wait for
      data with tags that $n$ generates.

\item Partitions that have \op{Recv} nodes that are connected in the previous
      step get a copy of this graph.

\end{itemize}
Deadness propagation and tag tracking, as outlined above, were the two most
important issues we had to tackle when extending TensorFlow's distributed
execution for tagged recursion.

\section{Evaluation} \label{evaluation}

In this section we evaluate our approach and implementation in terms of
performance. In particular, we first test our approach against some classic
microbenchmarks designed to stress recursive calls. Moreover, since TensorFlow
is primarily focused on machine learning tasks we test our approach in a
recursive neural network architecture.


\subsection{Functional Microbenchmarks}

\begin{table}
\centering
\begin{tabular}{lrrr}
  \toprule
    Function & Static (sec) & Dynamic (sec) & Speed-up \\
  \midrule
  {\tt fib(24)} & 0,812  & 0,990  & 18,00\%\\
  {\tt fib(25)} & 1,259  & 1,629  & 22,31\%\\
  {\tt fib(26)} & 2,127  & 2,572  & 17,31\%\\
  {\tt fib(27)} & 3,295  & 4,178  & 21,15\%\\
  {\tt fib(28)} & 5,338  & 6,739  & 20,79\%\\
  {\tt fib(29)} & 8,614  & 10,937 & 21,24\%\\
  {\tt fib(30)} & 13,968 & 17,801 & 21,53\%\\
  {\tt fib(31)} & 22,717 & 28,610 & 20,60\%\\
  {\tt fib(32)} & 37,446 & 46,855 & 20,08\%\\
  {\tt fib(33)} & 61,287 & 75,646 & 18,98\%\\
  \midrule
  {\tt ack(3,3)} & 0,089    &  0,118   &  24,76\% \\
  {\tt ack(3,4)} & 0,348    &  0,503   &  30,74\% \\
  {\tt ack(3,5)} & 1,511    &  2,095   &  27,88\% \\
  {\tt ack(3,6)} & 6,165    &  8,323   &  25,92\% \\
  {\tt ack(3,7)} & 25,489   &  33,147  &  23,10\% \\
  {\tt ack(3,8)} & 100,882  &  128,955 &  21,77\% \\
  \midrule
  {\tt tak(24,16,8)} &  18,881   &  16,075  &  -17,46\%\\
  {\tt tak(25,16,8)} &  29,545   &  25,394  &  -16,36\%\\
  {\tt tak(26,16,8)} &  45,218   &  38,722  &   -16,78\%\\
  {\tt tak(27,16,8)} &  67,757   &  57,876  &   -17,07\%\\
  {\tt tak(27,17,8)} &  188,506  & 159,778  &  -17,98\%\\
  \midrule
  {\tt primes(7500)}  & 13,502   &  13,458  & -0,33\%\\
  {\tt primes(8000)}  & 14,836   &  14,784  & -0,35\%\\
  {\tt primes(8500)}  & 16,268   &  16,076  & -1,19\%\\
  {\tt primes(9000)}  & 17,556   &  17,348  & -1,20\%\\
  {\tt primes(9500)}  & 18,868   &  18,912  &  0,23\%\\
  {\tt primes(10000)} & 20,636   &  20,378  & -1,27\%\\
  \bottomrule
\end{tabular}
\caption{Comparison of execution times (in seconds) in various invocations of well-known recursive function definitions. In particular, {\tt fib} computes the Fibonacci sequence, {\tt ack} is the Ackermann function, {\tt tak} is the Takeuchi function and {\tt primes} is the prime-counting function.}
\label{table:microbenchmarks}
\end{table}

We used a variety of recursive functions that due to their definitions in terms
of heavy use of recursion and demanding number of computations are regularly
used as benchmarks for recursion optimization tasks. We compare the execution
time of the tagged approach against the one that is already implemented in
TensorFlow (i.e., the expanding graph).

Table~\ref{table:microbenchmarks} presents the average execution times of
different execution of these microbenchmarks. The two implementations are
essentially comparable in the single-machine setting as expected. Both
implementations have their shortcomings in terms of imposed overhead in the
execution engine. On one hand, the dynamic implementation suffers from an
overhead due to the repetitive creation and initialization of the execution
environment. This overhead seems to be analogous to the number of invocations
and it is evidenced by the constant speed-up factor on the execution of the
function {\tt fib} (i.e., the recursive implementation of Fibonacci sequence).
On the other hand, the tagged implementation is slower in the {\tt tak} (i.e.,
the Takeuchi function). This is due to the inefficient manipulation and
management of the frames in the execution engine. In particular, there is an
overhead for concurrently accessing the global frame stack that limits the
parallelism. This phenomenon is more evident for functions with many arguments.
Recall that the frame stack was initially used by TensorFlow's engine to store
static scopes of while-loops. We believe that this inefficiency can be lifted by
carefully redesigning the execution engine

We generally believe that functions with definitions that comprise a large
number of operations tend to behave worse in terms of performance when executed
by the dynamic approach because of the repetitive copying of their large encoded
graphs at runtime whereas functions with smaller definitions do not suffer as
much by this particular overhead. However, a static approach may have the
potential to impose an even greater challenge to the current implementation of
TensorFlow, given that there still lies some space for further internal
optimizations regarding the management of frames and tags.


\subsection{Machine Learning Tasks}

\begin{table}
\centering
\begin{tabular}{lrr}
  \toprule
    Method & Training & Inference \\
  \midrule
  Unrolling & 0.99    & 1.74\\
  Iteration & 84.56   & 138.8\\
  Recursion & 143.33  & 161.29 \\
  \bottomrule
\end{tabular}
\caption{Comparison of three different implementations of a TreeRNN machine learning task. The measurements are throughput (instances per second) for both training and inference.}
\label{table:rnn}
\end{table}

\begin{table}
\centering
\begin{tabular}{lrr}
  \toprule
    Batch Size & Iteration & Recursion \\
  \midrule
  1  & 84.56   & 143.33 \\
  5  & 186.96  & 103.06 \\
  10 & 228.60  & 101.27 \\
  25 & 235.69  & 110.72 \\
  \bottomrule
\end{tabular}
\caption{Comparison of the throughput (instances per second) for two different
implementations of training a TreeRNN model with different batch sizes.}
\label{table:batches}
\end{table}

As a real-life example we experimented with the simplest model that belongs in
the greater family of Recursive Neural Networks~\cite{SocherPWCMNP13} also known
as TreeRNN. In particular, we consider the task of predicting the sentiment of
English phrases using a snapshot of the Stanford Sentiment Treebank dataset for
training our model. In this dataset every data instance is a fully labeled
parsed tree encoding the implicit tree-like structure of a certain sentence.
While this task is recursive in nature the training can proceed bottom-up and as
a result can be computed iteratively.

We compare the recursive implementation with two other implementations. First,
the most straightforward implementation is to statically unroll the computation
graph during its construction so that its form resembles the tree structure of
the given data instance. This method generates a separate dataflow graph per
input that will be executed only once throughout the overall training. The
second implementation is using an iterative mechanism as a means to process the
nodes of a particular tree in a bottom-up fashion.

Table~\ref{table:rnn} presents the performance of the three implementations of
TreeRNN. We measure the throughput (i.e., instances per second) for both
training and inference. We train on a set of 700 examples in total. As expected,
the static unrolling approach has the worst throughput of the three. The
iterative method performs also worse than the recursive. The parallelism in the
iterative method is limited since it processes each level of the tree
sequentially. On the other hand, the recursive method can process different
levels of the tree concurrently.

We also experimented with different batch sizes during training to observe how
this affects the throughput. Table~\ref{table:batches} compares the iterative
and recursive method under training sessions with different batch sizes. The
iterative method seems to benefit from batching and improves as batch size
increases. On the other hand, the recursive method does not seem to benefit from
the increase of the batch size. This is expected since each instance produces
different invocations of the operators. However, we believe that we can optimize
the execution of recursion in order to take into consideration the batch
execution of the same operator scheduled under different tags. This optimization
was out of scope of this work and left as future direction.

As an overall remark, the recursive method can express computations that can
make more dynamic decisions during runtime. We choose to evaluate on the
simplest recursive neural network in order to have the chance to compare with
different approaches. However, it should be noted that more dynamic neural
networks, such as Top-down TreeLSTM~\cite{tdtreelstm}, cannot be expressed
using iteration and therefore could not be used in this evaluation experiment.
We should note here that there is a recent recursive proposal based on the ideas
of expanding graphs~\cite{JeongJKYC18} (a more detailed discussion can be found
in Section~\ref{related-work}) that support automatic differentiation and could
be possibly used as a comparison on machine learning tasks. However, we could
not find any working implementation of that approach.

\section{Related Work} \label{related-work}
In this section we present a comparison of the proposed technique initially with
approaches that were developed in the early days of dataflow, and then with more
contemporary systems.
\subsection{Tagged Dataflow Systems}
The implementation of recursive functions using a tagged approach, has been
explored even from the early days of dataflow. In the Manchester prototype
dataflow computer~\cite{manchester-dataflow}, in order to implement first-order
recursive functions, tags are extended with an \emph{activation name} which is
used to distinguish different function invocations. In the MIT tagged-token
architecture~\cite{mit-dataflow}, a similar approach is used: each token is
tagged with a \emph{context identifier} that specifies the activation to which
the token belongs. In this way, tokens corresponding to different activations
may flow simultaneously through the dataflow graph, offering increased
parallelism. As it is noted in~\cite{mit-dataflow} the context identifier is the
dataflow analogue of the ``frame pointer'' in traditional activation records.
Both of the above works were groundbreaking and created the conceptual basis for
the computational paradigm of tagged dataflow. The present work differs from
both of the above techniques: we propose an algorithm for transforming any set
of dataflow graphs corresponding to recursive functions, into a single static
dataflow graph. Both~\cite{manchester-dataflow} and~\cite{mit-dataflow} describe
execution mechanisms for recursive functions, without explicitly describing an
algorithm for deriving a static dataflow graph. The static graphs we obtain are
of paramount importance in our more contemporary framework: they helped us in
extending the distributed computational model of TensorFlow to handle recursion,
and they guided us in obtaining an efficient procedure for automatic
differentiation. Both of these issues are extremely important for modern deep
learning applications.

In~\cite{yaghi}, a framework is developed for formalizing the implementation of
first-order recursive functions in a tagged manner. More specifically, it is
demonstrated that such an implementation can be achieved by transforming the
source recursive programs into simple \emph{intensional} definitions which can
then be executed in a demand-driven dataflow approach. This transformation
introduced in~\cite{yaghi}, is further examined and proven correct
in~\cite{jfp1}. Finally, in~\cite{jfp2} it is demonstrated that a large class of
\emph{higher-order} recursive functions can also be implemented in a tagged
manner. These approaches have been later used in order to derive an
implementation of a general-purpose lazy functional
language~\cite{CGPR08,FPR13}. In these approaches the tags are also represented
as a global stack similarly to the implementation presented in this paper. The
major difference, however, is that the execution proceeds in a demand-driven
instead of a data-driven manner. The graph is not explicitly constructed during
execution and no offloading to specialized computational units was considered.
Moreover, none of these implementations address distributed execution nor
automatic differentiation which is essential for machine learning applications.
However, it would be interesting to investigate how the technique developed
in~\cite{jfp1} can be adapted to data-driven execution and then to examine how
the resulting transformation is semantically related to the graphs that result
from the proposed transformation algorithm.

Naiad~\cite{naiad}, a more recent distributed dataflow system focused on data
processing, also makes heavy use of tags. Naiad implements the differential
dataflow model~\cite{diffdataflow} that employs tags to support
incremental and iterative computations. However, Naiad does not support
recursive dataflows.

\subsection{Machine Learning Dataflow Systems}

Machine learning frameworks that follow the ``in-graph'' approach, such as
TensorFlow~\cite{tensorflow} and Theano~\cite{theano}, encode dynamic
control-flow decisions in the graph. However, user-defined functions are not
fully supported by these two systems. More specifically, both systems support
non-recursive function definitions. TensorFlow applies an optimization that
inlines the body of a non-recursive function in the main dataflow graph. On the
other hand, Theano does not support recursive definitions while TensorFlow
supports them only partially without supporting automatic differentiation on
them. TensorFlow executes a recursive function by dynamically expanding the
graph at runtime when a recursive call is encountered. In order to support
algorithms that recursively traverse tree-like data structures, a
transformation~\cite{LooksHHN17} must be applied that transforms recursion to
iteration that TensorFlow supports. Transforming recursion to efficient
iteration is not generally a straightforward task.

A recent related work that tries to improve recursion support in
TensorFlow~\cite{JeongJKYC18} employs the same approach of the dynamically
expanding graph to execute recursive functions and defines a recursive automatic
differentiation technique of such functions. However, that work considers only a
single-machine execution. The main disadvantage of this particular approach is
that a dynamic graph cannot contain the subgraphs that correspond to the
recursive callee functions, thus, they cannot be distributed without central
coordination. Every special node representing the calling of a function resides
on one partition and initiates the execution of the function by just one worker.
That means, that the execution of functions cannot be easily shared amongst
multiple machines.

Other machine learning frameworks, such as Torch~\cite{torch},
PyTorch~\cite{pytorch}, DyNet~\cite{dynet} and Chainer~\cite{chainer} follow the
``out-of-graph'' approach, that is they rely on the host language capabilities to
encode control-flow decisions. This lacks efficiency because the ability to
perform optimizations on the overall graphs is significantly limited.

\subsection{Other Dataflow Systems}

CIEL~\cite{ciel} represents a program as an unstructured dynamic task graph in
which tasks can tail-recursively spawn other tasks. CIEL supports distributed
recursive algorithms by transforming them into a continuation passing style.
Contrary to the current work, CIEL makes use of a master node that maintains a
list of pending tasks and acts as a task dispatcher to the workers. The major
drawback of unstructured dynamic graphs is that they are much more challenging
to optimize holistically.

\section{Conclusions and Future Work} \label{conclusions}

In this paper we proposed an algorithm for transforming a set of dynamic (ie.,
expandable) dataflow graphs into a single static (ie., non-expandable) graph.
Our technique can be extended to support automatic differentiation, something
really important for machine learning applications. All the proposed ideas have
been implemented in the TensorFlow framework.

We believe that \emph{tagging} is a very promising direction that current
implementations of dataflow systems should further exploit. Tags offer increased
parallelism and may prove to offer viable solutions in cases where current
technologies appear to bottleneck. As possible future directions, we would like
to investigate the possibility of implementing \emph{higher-order} recursive
functions under the TensorFlow framework. In~\cite{jfp2} it was suggested that
this can be achieved by using more complicated tags, an idea that is worth
further consideration. Additionally, we would like to investigate the
possibility of implementing user-defined data structures in a purely tagged
manner. Finally, we would like to investigate the relationship between
\emph{multidimensional programming}~\cite{multidimensional} (a form of
programming that was proposed more than 25 years ago as an extension of
classical dataflow programming) and the use of tensors in modern dataflow
systems. More specifically, tensors appear to be closely related to the
multidimensional streams introduced in~\cite{multidimensional}, and therefore
the techniques introduced in~\cite{multidimensional} may be applicable to modern
tensor-based systems. In conclusion, we believe that tagged dataflow is a very
interesting computational paradigm, and its resurgence through the  appearance
of modern dataflow systems, may lead to exciting new developments both in
foundational as-well-as practical issues.


\bibliography{recfun-in-tf}

\end{document}